# Chapter 1

# STELLAR AND DYNAMICAL EVOLUTION WITHIN TRIPLE STARS

PETER P. EGGLETON, LUDMILA G. KISELEVA
*Institute of Astronomy*
*Madingley Rd, Cambridge CB3 0HA, United Kingdom*
`ppe@ast.cam.ac.uk, lgk@ast.cam.ac.uk`

## Abstract

About 5-15% of stellar systems are at least triple. About 1% of systems with a primary of $\gtrsim 1\,M_\odot$ are triple with a *longer* period that is less than 30y, and so may in principle be capable of Roche-lobe overflow in both the inner and the outer orbits, at different times. We discuss possible evolutionary paths for these systems, some of which may lead to objects that are difficult to understand in the context of purely binary evolution. An example is OW Gem, a binary containing two supergiants (spectral types F and G) with masses that difffer by a factor of 1.5. There is also a triple-star pathway which could lead rather naturally to low-mass X-ray binaries; whereas binary pathways often appear rather contrived. We also discuss some dynamical processes involved in the 3-body problem. A number of triple stars are found in clusters. Similar systems can be created by gravitational capture during N-body simulations of Galactic clusters, especially if there is an assumed primordial binary population. We discuss the properties of these triples, and note that many can be quite long-lived.

## 1.1 Introduction

Among field stars, triple and higher multiple systems are reasonably common. Of the order of 5-15% of all stellar systems are at least triple. There is marginal evidence that the degree of multiplicity of systems increases somewhat with primary mass. Among the 50 nearest systems (van de Kamp 1971, Henry & McCarthy 1990), mostly G/K/M dwarfs, there appear to be 33 singletons, 13 binaries and 4 triples, and among the 164 nearest solar-type dwarfs Duquennoy & Mayor (1991) state that there are 93 singletons, 62 binaries, 7 triples and 2 quadruples. These somewhat similar samples show that ∼ 8% and 6% respectively of fairly low-mass systems have multiplicity higher than two. On the other hand, among the 50 brightest systems (Hoffleit & Jaschek 1983, Batten, Fletcher & McCarthy 1989), mostly B/A dwarfs and G/K giants, there appear to be 27 singletons, 15 binaries, 3 triples, 4 quadruples and one sextuple. The proportion of triples and higher multiples is ∼ 16%. Perhaps there is a slight trend here, towards higher multiplicity at higher mass, although the statistical significance is not high.

Virtually all observed multiples are 'hierarchical' (Evans 1968): a close binary has a distant companion, which may itself be a close binary; or a close binary with distant companion has a considerably more distant further companion still. The well-known sextuple $\alpha$ Gem is a wide hierarchical triple in which each of the 3 visual components turns out to be a short-period spectroscopic binary. For the most part, we will only consider systems with multiplicity 3 rather than 4, 5 or 6, and in that case we can conveniently talk about an 'inner' binary, often with period a few days, and an 'outer' binary, one component of which is the inner binary, with a period of typically a few years. Of course most known triples, like most known binaries, are too wide for there to be significant binary interaction in the form of Roche-lobe overflow (RLOF) etc., but there is a proportion of triples in which *both* orbits are sufficiently small that RLOF might take place in the outer as well as the inner orbit, at different evolutionary stages. We shall refer to such systems as 'doubly interesting' triples.

It is not easy to pin down the proportion of stellar systems which are doubly interesting. Among the ∼ 5000 stars brighter than $V = 6.0$ are at least 30 where the *outer* period is ≲ 30y, which I shall take as the maximum period for RLOF, following Plavec (1968) and Paczyński (1971) in their discussion of binaries. This amounts to 0.6%. But there are several selection effects operating against the discovery of outer orbits with periods as short as months to a few years, and also of inner orbits with periods this long, and so we suspect that 0.6% is very much a lower limit. Dynamical stability requires that the outer period be longer than the inner period by a factor of ∼ 3 − 6 (if both orbits are nearly circular; Kiseleva, Eggleton & Anosova 1994), or more generally that outer periastron be larger than inner apastron by a factor of ∼ 12.3 − 16 (Harrington 1975, Eggleton & Kiseleva 1995). These factors assume that all three bodies are of comparable mass, i.e. within a factor of 100 of each other. For hierarchical systems of more extreme mass ratio, such as a star and 2 or 3 planets, the periods can be closer together.

In sect. 1.2 we consider some of the effects that stellar evolution can be expected to have in such doubly-interesting systems. Some of these effects may lead, through coalescence of two components of the system, to binaries with properties that would be difficult to account for within the context of purely binary evolution. In sect. 1.3 we consider aspects of the gravitational dynamics of triples, especially those in which the three components are sufficiently close that the hierarchical nature of the initial system can be expected to break down. In sect. 1.4 we consider the formation of triples within an N-body simulation of a Galactic cluster. Especially if there is a primordial distribution of binaries, hierarchical triples that persist for a considerable time can be formed by purely gravitational encounters.

## 1.2 Stellar Evolutionary Effects

The following examples, taken from the comprehensive discussion of Fekel (1981), illustrate the kind of doubly-interesting systems in which internal stellar evolution can play a role.

$\beta$ Per   ((K0-3IV + B8V; 0.8 + 3.7 $M_\odot$; SD, 2.87d) + F1V; 4.5 + 1.7 $M_\odot$; 1.86y, $e$=0.23)

$\beta$ Cap   ((B8V + ?; 3.3 + 0.9 $M_\odot$; 8.68d) + K0II-III; 4.2 + 3.7 $M_\odot$; 3.76y, $e$=0.42)

$\lambda$ Tau   ((A4IV + B3V; 1.9 + 7.2 $M_\odot$, SD, 3.97d) + ?; 9.1 + 0.7 $M_\odot$; 0.09y=33d, $e \sim 0.15$)

Where the eccentricity is not given, it is zero to observational accuracy; SD stands for semidetached. Note that Lestrade et al. (1993) have shown by VLBI that the two orbits of $\beta$ Per above, while both inclined at $\sim 90°$ to the line of sight, are inclined at $\sim 100°$ to each other, i.e. far from being coplanar, they are in fact slightly retrograde. This strikes a cautionary note, since it is usually assumed that such orbits are approximately coplanar. In the case of $\lambda$ Tau, however, there is good evidence that the two orbits are nearly coplanar, from the fact that the outer orbit would otherwise, through precession, cause the eclipses in the inner orbit to vary slowly in time (Söderhjelm 1975, Fekel & Tomkin 1982). But even here we might wonder whether the relative inclination is near 180° rather than 0°.

To the above 3 examples we add one trebly-interesting quadruple system – in fact, one of only two that we are aware of, although we imagine that they may not be all that rare:

$\mu$ Ori   ((A7m + ?; 1.8 + 1 $M_\odot$; 4.45d) + (F3V + F3V, 1.4 + 1.4 $M_\odot$; 4.78d); 2.8 + 2.8 $M_\odot$; 18.8y, $e$=0.76)

The notation for the four multiples above, involving nested parentheses, follows loosely the suggestion of Evans (1977). The inner parentheses contain a description of the components and the orbit of the inner binary or binaries, and the outer parentheses relate to the components of the outer orbit. The masses given above, taken from Fekel (1981), contain in some cases an element of inference: not all of these systems have yielded so much information that all three (or four) masses are unambiguously determined.

Several systems are known with deeper levels of hierarchy than two, although none, so far as we are aware, has three levels of hierarchy with the outermost of the three or more orbits having period $\lesssim 30$y. Such systems might be especially difficult to recognise, since the middle orbit could fall uncomfortably between the range of spectroscopic orbits at small separations, and of visual orbits at large separations. Note that the above four systems, as well as four others ($\xi$ Tau, $\eta$ Ori, $\kappa$ Peg and p Vel) are all among the brightest 500 stars, a set which may be reasonably representative of stars with masses $\gtrsim 1\,M_\odot$; thus it seems possible that doubly or trebly interesting systems may represent $\sim 1\%$ of such systems, although of course multiples are bound to be somewhat over-represented in a magnitude-limited sample. It would not be surprising if a further one or two systems of these 500 are similarly multiple, given particularly the difficulty of recognising small third bodies such as M dwarfs at separations of a few AU from B/A companions.

As an aside, and at the risk of annoying our observational colleagues, we cannot refrain from saying that occasionally the biggest difficulty that theorists like ourselves have in reading a paper on some new hierarchical system is disentangling which component is which, and how the orbits are thought to be nested. Partly this is because the words 'primary' and 'secondary' can mean different things in different contexts, and partly because words like 'star' or 'component' can sometimes mean a combination of more than one stellar entity, again depending on context. We would like to suggest that every paper on a new multiple give, at or very near the beginning, a description involving nested parentheses in the style of Evans (1977) – as we have written above for $\beta$ Per etc., although in many cases only a mass function, rather than two values of $m \sin^3 i$, may be available, or an angular separation rather than an orbital period. Such a notation makes it very much easier to follow the remaining argument. In some cases, of course, there is genuine ambiguity about the way in which the components are grouped in the hierarchy, because observations cannot necessarily distinguish all possible configurations; but even in this case, it would be very helpful to see the various options that are consistent with the observations written in the above very compact and yet very explicit way.

Non-hierarchical multiples are known, such as the Trapezium in Orion's sword, but they are fairly rare. This is not surprising in view of the fact that they should be dynamically unstable on the timescale of several crossing-times (see e.g. Anosova & Orlov 1994 and references therein). Such multiples may represent a sub-clumping in a young cluster, and it is likely that after one or two stars have been expelled, the remainder may settle down in a long-lived hierarchical system. In sect. 1.4 we consider how it may be possible to distinguish, in the context of an N-body simulation, between short-lived, generally non-hierarchical, and long-lived hierarchical systems that can form (and also dissolve) by purely gravitational encounters.

It is not likely that gravitational encounters alone can lead to the formation of such multiples in the general Galactic field as the four listed above. In the context of binaries alone, it has proved very difficult to account for systems with periods of only days, of which there are five in the above four multiples. Clarke & Pringle (1991) considered whether the dissipative effect of collisions between discs around protostars might allow a triple, in an initial equilateral-triangular configuration which rotates at a rate well below centrifugal-gravitational balance, to end up as a stable hierarchical system. Although, in 50 simulations with somewhat different randomly-distributed initial parameters, they found that 12 potentially stable hierarchical triples were formed, the smallest inner semi-major axis was $\sim 10\,\mathrm{AU}$, and the outer semi-major axes were $\sim 2000 - 50\,000\,\mathrm{AU}$. Some of their triples were not evolved far enough to be sure of the outcome, but they estimated that these systems would most probably be broken up in the longer term.

We suggest that a possible mechanism may be rather similar to the common-envelope (CE) evolution of Paczyński (1976). This process is almost certainly responsible for the fact that some initially wide binaries ($P \sim 1 - 10\,\mathrm{y}$) must, after one component reaches the AGB, evolve into close binaries with $P \sim 0.1 - 1\,\mathrm{d}$. Such binaries are found in the centres of planetary nebulae, *e.g.* UU Sge (SDO + F:V, $0.63 + 0.29\,M_\odot$; 0.465d; Bond 1976, Pollaco & Bell 1993) in the PN Abell 63. If one protostar is surrounded by a roughly spherical cocoon with mass and dimensions comparable to an AGB star envelope, and if a companion on an only moderately elliptical orbit were to become 'trapped' in this cocoon, the companion might be obliged to spiral in while transferring angular momentum to the cocoon, but without dissipating the cocoon so quickly that the interaction ceases too soon. Such a mechanism is very tentative, since it is difficult, for example, to see what would support such a cocoon in the absence of the very high internal nuclear luminosity that supports an AGB star's envelope. In any event, the formation of truly close binaries, with periods of a few days, remains an important but elusive part of astrophysics.

The question which we wish particularly to address in this Section is not the formation, however, but rather the subsequent evolution: how might internal evolution affect 'doubly interesting' triples? In particular, are there any evolutionary channels open to triples which are not open to mere binaries? We may attempt to consider what the future holds for systems such as the four above, and also for some other known systems. And we may also speculate on systems which, while somewhat similar to those above, might not yet have been observed because of the observational difficulties: for example, a system like $\lambda$ Tau but with the outer orbit being $\sim 300$d rather than 33d.

We can start by distinguishing broadly between (a) triples in which the distant component ($*3$) is the most massive of the three, and (b) triples in which $*1$ ('star 1', by convention here the originally more massive of the two in the inner binary) is the most massive of the three. Throughout this paper the identification of $*1$ and $*2$ is based on *initial masses*, and not on current masses or luminosities or temperatures, so that the labels do not change in the course of evolution even although the masses and other properties can change by large amounts in either direction. The reader may note that this convention was followed in the double-parenthetical descriptions of the systems at the beginning of this Section: in the two which contain Algols it is evidently the loser which was *initially* more massive.

In case (a), of which $\beta$ Cap above is an example, we expect RLOF from $*3$ to ($*1+*2$) before either $*1$ or $*2$ can evolve significantly. Such RLOF is more likely to be stable, i.e. to proceed on only a thermal or nuclear timescale rather than on a hydrodynamical timescale, than is the case in a binary which is comparably wide. This is because the stability depends on the mass ratio in the outer orbit, which will generally be reduced by virtue of the close binarity of the companion. In most *observed* triples, $m_3 < m_1 + m_2$, even though in case (a) $m_3 > m_1$. To be confident of stability, we would like $m_3 \lesssim 0.7(m_1 + m_2)$, since only for such mass ratios will the loser's Roche lobe, under conservative assumptions, expand faster than the loser itself on its Hayashi track. This condition is not satisfied by $\beta$ Cap; but the outer mass ratio of 0.88 is at least substantially nearer to stability than the value of 1.12 that would apply if $*2$ were missing. Thus instead of a major episode of common-envelope evolution, as expected in a normal Late Case C binary (Paczyński 1976), we should rather expect a minor episode that might only shrink the outer orbit by a modest factor. There are several triples in which the outer mass ratio is substantially smaller still, approaching the limit of 0.5 for three equal stars (a surprisingly common situation, no doubt partly because of selection effects); such systems should certainly give reasonably stable RLOF.

If the gainer in a wide binary is itself a close binary, then the accretion process is likely to be strongly modified (Eggleton & Pringle 1985). Possibly no accretion will occur at all. The action of the close binary may force the material to accumulate in a disc with a fairly definite inner edge at a few times the inner binary's separation, and then torque from the close binary on the disc may ultimately dissipate the disc (an 'excretion disc'), though probably on a fairly slow timescale. Such a process was suggested by Eggleton & Pringle (1985) as a possible explanation of the unusual system $\epsilon$ Aur (F0Ia + IR disc; 9890d; Huang 1965, Wilson 1971, Lissauer & Backman 1984, Van Hamme & Wilson 1986). The F0Ia component is presently quite far inside its Roche lobe, but it may be that this component was an M supergiant a few thousand years ago, and is either proceeding towards a white dwarf or else is in a temporary contraction due to a shell flash. In a purely binary picture it is difficult to see why the disc is still there; but if in the centre of the disc there is a hole (Wilson 1971, Van Hamme & Wilson 1986), and if in that hole there is a binary, the situation is less puzzling. This possibility can also explain why no stellar companion has ever been detected – if the companion is a binary of say two A dwarfs, it would be quite undetectable next to an F0Ia supergiant, whereas if the companion had the same mass as the supposed close binary it would probably be detectable as a mid-B dwarf.

The observational fact mentioned above that almost always $m_3 < m_1 + m_2$ may partly be another selection effect; for if the combined mass of the inner binary is less than the third mass, the two individual components will be less massive still, and so the inner binary may be very faint. Griffin (1985, 1986) has found some G/K giant single-lined spectroscopic binaries in which the mass function is sufficiently large that one would expect the companion to be detectable. He interprets the fact that it is not detected as an indication that the companion may be a short-period binary. Thus there may well be systems where a common-envelope phase takes place in which a short-period binary spirals into a red supergiant envelope. We suspect, however, that this is never as extreme a process as in the kind of system that leads to $\sim 0.5$d systems such as UU Sge above.

Conceivably the quite unusually close triple $\lambda$ Tau might itself be the result of spiral-in evolution in a system which was initially much wider, provided that the unseen $*3$ of $\sim 0.7\,M_\odot$ (Fekel & Tomkin 1982) is a white dwarf rather than an MS star. There is no observational knowledge to contradict this, although at the same time there is nothing to support it except the fact that the outer period is so unusually short as to be in the regime where one thinks about spiralling-in as a mechanism even for binaries let alone triples. If the two components of the Algol pair were initially $\sim 4.5\,M_\odot$ each, and if $*3$ was perhaps $7\,M_\odot$, then such an evolution might have occurred. The WD probably should be $0.8 - 0.9\,M_\odot$, but this may not be inconsistent with the observed value, which has substantial error bars.

Whatever the present evolutionary state of $*3$ in this system, the future evolution is likely to see the Algol pair get wider, as further mass is transferred, until at a period of $7 - 10$d (see sect. 1.3 for more details) it is sufficiently wide that dynamical interaction with $*3$ will result in the ejection of $*3$ from the system (Bailyn & Eggleton 1983, Kiseleva, Eggleton & Anosova 1994, Kiseleva, Eggleton & Orlov 1994). This seems likely to be the general fate of low-mass close companions to Algols, if the initial period ratio is small; but low-mass companions that are further away will not be ejected, and may instead get caught up in a spiral-in phase if the Algol pair later coalesces into a single red giant (see below), which evolves to an AGB state.

Before considering case (b) as defined above, let us consider an intermediate case, say (ab), where $m_1 \sim m_3$ to within a few per cent. In such a case *1 and *3 evolve at much the same rate. The result will be a system containing an Algol like

DL Vir   ((? + A3V, 1.4 + 2.5 $M_\odot$; SD, 1.32d) + G8III, 3.9 + 2.2 $M_\odot$; 6.2y, $e=0.44$)

where the masses are again taken from Fekel (1981). The close companion to the A3 star is presumably also a giant, or at any rate a subgiant, and we can infer that its initial mass was much the same as *3, i.e. the close pair started with masses 2.2 + 1.7 $M_\odot$. This is a rare case in which the *initial* masses of an Algol pair can be determined with rather little inference. It will be very important to have high-quality data for this triple, so that models of RLOF, both conservative and non-conservative, can be tested more thoroughly than is possible for most Algols. Chambliss (1992) has catalogued 79 eclipsing systems with known companions, 29 of these systems having known outer as well as inner orbits (although a few of the former are not yet well-determined). Of the 29, 9 are probably SD systems (Algols), but DL Vir appears to be the only one in which *3 as well as *1 is evolved to a giant or subgiant.

Turning to case (b), we expect RLOF first in the inner binary. Provided that the initial mass ratio is not very extreme, and that the orbital period is not too short, the inner system can avoid either of the two types of contact that are possible (Eggleton 1995) and so evolve as a semidetached system. Although the long-term fate of such Algols is not yet clear, mainly because there is scope for both angular-momentum loss (AML) and mass loss (ML), we can anticipate something like:

(MS + MS, D) → (RG + MS, SD) → (WD + MS, D) → (WD + RG, SD) → CE ...

where MS means a main-sequence component, RG means a red giant, WD means a white dwarf, and D means detached. The outcome of the CE phase can be expected to be either a single, coalesced star, or else a close binary of two low-mass WDs. The CE phase seems almost inevitable, since the mass ratio at the WD + RG phase is likely to be very high. For example, in AS Eri (K0III + A3V, 0.2 + 2.0 $M_\odot$; 2.66d; Popper 1980) the RG must be close to becoming a WD, and the mass ratio is already 1:10. AS Eri, incidentally, is clear evidence that AML at least is important (Refsdal, Roth & Weigert 1974), since the present angular momentum is much too low for the system to have avoided implausibly deep contact at the earlier stage when the masses of the two components were equal. The reverse RLOF which AS Eri faces in the future when the A3 star in turn evolves must surely be very dramatic. We believe that at the short period of this system the most likely outcome is CE evolution followed by coalescence, although for Algols of somewhat longer period the CE evolution might rather lead to a very close WD + WD pair.

If we return to DL Vir, the three components' masses might evolve roughly thus:

$$((2.2 \text{ MS} + 1.7 \text{ MS}) + 2.2 \text{ MS}), \text{ initially} \rightarrow ((1.4 \text{ RG} + 2.5 \text{ MS}) + 2.2 \text{ RG}), \text{ now}$$
$$\rightarrow ((0.2 \text{ WD} + 3.7 \text{ MS}) + 2.2 \text{ RG})$$
$$\rightarrow ((0.2 \text{ WD} + 3.7 \text{ RG}) + 2.2 \text{ RG})$$
$$\rightarrow (3.9 \text{ RG} + 2.2 \text{ RG}),$$

where we have made the supposition that that the distant RG will not complete the whole of its evolution, including a fairly long-lived core-helium-burning phase, in the time it takes for (i) the other RG to lose all its envelope, and (ii) the massive MS star so formed to evolve to an RG itself. This may not be entirely realistic, but we need only suppose that *3 might be slightly less massive originally, and so reach its RG stage a little later. Then the final outcome of the above sequence is a *binary* rather than a triple system, in which there are two red giants of very different masses, one nearly twice the mass of the other. Compare this with the rather remarkable binary OW Gem (F2Ib-II + G8IIb, $6.0 + 4.0\, M_\odot$; 3.45y, $e = 0.52$; Griffin & Duquennoy 1993). The masses here are very well determined, the system being both double-lined and doubly-eclipsing. This system is difficult, one might say impossible, to achieve on the lines of conventional evolution of two stars that have never interacted; and it is likely that the two stars have never interacted because the eccentricity remains quite high. It could, however, be the legitimate outcome of a system that was initially triple – somewhat like DL Vir, though with higher masses for all three components.

Wider Algols than AS Eri or DL Vir might avoid coalescence, and instead end up as close pairs of low-mass WDs after CE evolution. A system that might be nearly there is V1379 Aql (SDB + K0III, $0.31 + 2.34\, M_\odot$; 20.66d; Jeffery, Simon & Lloyd Evans 1992). Since both hot subdwarfs and red giants are relatively short-lived phases of evolution compared with the MS stage, one can probably infer for this system that it started with fairly nearly equal masses, so that in its previous Algol stage *both* components were red giants simultaneously, as for example in the 'cool Algols' RT Lac, RV Lib, AR Mon and RZ Cnc (Popper 1980, his Table 13), whose periods range from 5 to 22d. We can also infer that when the present RG fills its Roche lobe its WD core will have a mass that is much the same, but a little larger (say $0.35\, M_\odot$), than the present SDB component – slightly larger, because although its Roche lobe is $\sim 2.5$ times larger, thanks to the mass ratio of $\sim 8$, the radius of an RG is quite sensitive to core mass and not very sensitive to anything else. We can only speculate at present on the extent to which the period is reduced by the impending CE phase: Marsh, Dhillon & Duck (1995) have discovered a number of WD + WD binaries with comparably low masses, and periods in the range 0.145 - 4.87d.

We should note here that in order to obtain such cool Algols as those listed above we must already make a modification to classical 'conservative' binary evolution. These Algols will all have been formed in Late Case B, where the loser has a deeply convective envelope by the time it fills its Roche lobe. Thus they should be expected, as for Late Case C, to go straight into CE evolution and not into a reasonably steady mass-transferring state as observed. A possible answer to this (Tout & Eggleton 1988) is that shortly before RLOF, when the more evolved star is a giant with a radius of $\gtrsim 30\%$ of its lobe radius, the enhanced dynamo activity that appears to be characteristic of these systems (RS CVns, Hall 1976) leads to enhanced mass loss and alters the mass ratio so that the system is more stable once RLOF begins. Z Her (K0IV + F5IV; 1.3 + 1.6 $M_\odot$; 3.99d, Popper 1988) appears to be heading along this route: the K0IV is still not filling its Roche lobe, and yet it is substantially less massive than its companion. Such a mechanism appears to be necessary prior to the first RLOF (from ∗1 to ∗2), in order to create a normal Algol from a detached binary that started perhaps with a mass ratio not very different from unity, but it is unlikely to be so strong that it can prevent a CE phase at the *second* RLOF (from ∗2 to ∗1), even although RS CVn-like behaviour can be expected, and is indeed observed, in V1379 Aql above and similar systems.

If close WD + WD binaries are typical products of Algol evolution, at least for wider Algols, then there are several triples already known in which there is the interesting prospect that a WD + WD binary will plunge into the extended red-supergiant envelope of a distant companion. We hesitate to speculate on the outcome, but it may well be somewhat easier in this scenario than in the conventional two-body scenario to accumulate enough mass of degenerate material to violate the Chandrasekhar limit. A close pair of $\sim 0.4\,M_\odot$ each, with a third WD of say $0.7\,M_\odot$, should do. The close pair would have to be the remains of a post-Algol of somewhat greater separation than V1379 Aql, but the cool Algol AR Mon (K3III + K0III, 0.8 + 2.7 $M_\odot$; 21.2d, Popper 1980) should double or treble its period before its loser becomes a WD of $\sim 0.4\,M_\odot$ (always supposing AML is not too effective). Thus a triple containing AR Mon as its inner pair, and with an outer period of a few years, might be a more natural precursor to a Type Ia SN than something which is merely binary.

If the inner binary is somewhat wider still, and headed for Case C RLOF, then it could evolve into a CV. There is one known 'doubly-interesting' triple which contains a CV (Reimers, Griffin & Brown 1988):

$$4\ \text{Dra}\quad ((\text{WD} + ?;\ 0.16\text{d}) + \text{M3III};\ 4.7\text{y},\ e{=}0.3).$$

In fact the outer orbit in this system is uncomfortably small for a system whose *inner* orbit is supposed to have reached Case C RLOF, although such evolution is not quite impossible (Eggleton, Bailyn & Tout 1989). We wonder, though, whether the AM Her-like CV component might actually be something rather different: a WD + WD binary that is accreting wind from the M giant, rather than a conventional CV. Such a possibility would allow the precursor inner binary to have come from a fairly substantial range of shorter periods, rather than from more-or-less the maximum period consistent with dynamical stability of the triple, which would be necessary for Case C.

We conclude this section with a brief look at the possibilities within more massive binaries, those in which one component at least is above $\sim 8\,M_\odot$, and is therefore likely to produce a neutron star (NS) rather than a WD. In fact $\lambda$ Tau above may just scrape in, since ∗2 has already reached 7.2 $M_\odot$ as a result of RLOF and is likely to exceed 8 $M_\odot$ in the future; but by that time ∗3 will probably have already been ejected by dynamical interaction. 'Doubly-interesting' massive triples are likely to suffer from the selection effect that, being typically more distant, a visual outer orbit will be harder to recognise. However, two examples are

η Ori    ((B1V + B3V, 15 + 12 $M_\odot$; 7.98d)  + B1V; 27 + 14 $M_\odot$; 9.5y, 0.044″, $e$=0.43)

VV Ori    ((B1V + B5V, 10.8 + 4.5 $M_\odot$; 1.49d)  + A3:, 15.3 + 2.3: $M_\odot$; 0.33y, $e$=0.3),

along with the only 'trebly interesting' massive system we are aware of:

QZ Car ((O9.7Ib + B2V:; 40: + 9: $M_\odot$; 20.7d, $e$=0.34)  + (O9V + B0Ib; 28 + 17 $M_\odot$; 6.00d);

49: + 45 $M_\odot$; $\lesssim$ 25.4y, $\lesssim$ 0.012″).

The data are from Fekel (1981), Chambliss (1992) and Popper (1993). The system η Ori contains at least one more component, but sufficiently distant to be 'uninteresting' in the context of potential RLOF. Note the very short outer period of VV Ori; as for λ Tau, ∗3 is recognised mainly by its influence on the motion of the centre of mass of the inner pair, although in VV Ori ∗3 is also recognised by its 'third light' contribution to the light curve.

Since a supernova explosion (SNEX) in a binary typically disrupts the binary, it probaby also typically disrupts triples, but some binaries are certainly left intact (though altered) and so some triples might also. Any of the systems above might later contain a high-mass X-ray binary (HMXB). In that case, an interesting possibility arises which is not possible in the context of purely binary evolution.

Although the long-term fate of HMXBs is by no means clear, it is likely that those of the shortest periods, say $P \lesssim 4d$, end up as coalesced stars, Thorne-Żytkow objects (TŻOs, Thorne & Żytkow 1977, Cannon 1993), via a CE phase, much as we expect the closer post-Algols to end up as coalesced RGs. Wider HMXBs can be expected, at least in some cases, to end up as double-NS binaries; but if the massive component of an HMXB is expected to fill its Roche lobe even before, or just after, the end of its MS evolution then it is likely that the NS – or perhaps a black hole (BH) – will plunge right into the centre of the massive companion. This should cause the companion to swell up to extreme red-supergiant proportions. If there is a ∗3 with (i) fairly low mass, and (ii) separation of up to a few (perhaps $\lesssim 20$) AU, then there will be a second CE phase, whose outcome (Eggleton & Verbunt 1986) could be a low-mass X-ray binary (LMXB) – in much the same way that an AGB star with a low-mass companion is thought to evolve into a CV. VV Ori (above) might therefore be a quite reasonable progenitor for a system like Her X-1 (Middleditch & Nelson 1976), and other more typical LMXBs with lower-mass companions; although we might prefer ∗3 in VV Ori to be somewhat further away. One advantage of such a scenario is that the low-mass ∗3 can be quite distant originally from the SNEX that forms the NS or BH, and thus is spared the risk of being entirely demolished. Such a risk is present in most purely binary scenarios, where the binary is assumed to have already shrunk drastically, thanks to CE evolution, *before* the SNEX that creates the NS or BH.

## 1.3   Orbital Parameters of Stable Hierarchical Triple Stars

It is not easy to be confident, for a particular set of three particles with given instantaneous positions and motions, whether they form a hierarchical system or not, and (if they do) whether it is stable or not. We have given (Eggleton & Kiseleva 1995) a fairly general and simple criterion involving the two mass ratios, and either the the eccentricities and the ratio of periods, or alternatively (and equivalently) the ratio of outer periastron distance to inner apastron distance. The effect of varying orbital inclination is relatively minor: it does not usually affect the critical period ratio by more than ~ 20%, and this is about the level of accuracy that the criterion aims for. This criterion comes from a considerable number of numerical simulations (Kiseleva, Eggleton & Orlov 1994), rather than from analytic insight. It appears to be more general and precise than the Harrington criterion (Harrington 1975, 1977) which involves similar parameters of triple stars.

In dynamically stable hierarchical systems, even small changes of such orbital parameters as eccentricity and semi-major axis due to perturbations by the distant component(s) may significantly affect the internal processes in close binaries (e.g. the mass transfer rate in semi-detached systems). The distant component is always pumping an eccentricity into the binary on the orbital (or shorter) time scale, even if at some stage the binary had zero instantenious eccentricity. In order to estimate the average value of eccentricity of both the inner and the outer binaries in stable hierarchical triples, we considered a wide range of both mass ratios, expressed by values $\alpha$ and $\beta$, where $\alpha \equiv \log_{10}(m_1/m_2)$ and $\beta \equiv \log_{10}\frac{m_1+m_2}{m_3}$, and of initial period ratio $X_0 \equiv P_{\text{out}}/P_{\text{in}}$ at time $t = 0$. In fig. 1.1 we show means over the time for which we computed the evolution ($0 < t < T \equiv 100 P_{\text{out}}$) of the values of inner and outer eccentricity as functions of $X_0$ for two of our simulations with different ($\alpha$, $\beta$) pairs. Note that for coplanar stable hierarchical systems with initially circular orbits the time of integration $T$ is not important, because both eccentricities show no secular changes; this was recently shown analytically by Heggie (private communication). He also showed that secular or more precisely long-term periodic changes may be very important for non-coplanar orbits (we will discuss the impact of this conclusion on triple systems formed in clusters in sect. 1.4). Also there are short-term fluctuations (with period equal to the inner or outer orbital period) of inner and outer orbital parameters (eccentricities and semi-major axes) around their average values. In the course of these fluctuations instantaneous values of both $e_{\text{in}}$ and $e_{\text{out}}$ go very close to zero periodically but their mean values are always $> 0$. In fig. 1.2 we show the behaviour in time of $e_{\text{in}}$ and $e_{\text{out}}$ over 40 inner orbits of a triple with three equal masses and stable period ratio.

We have found (Kiseleva, Eggleton & Anosova 1994) that the variation of $e_{\text{in}}$ with $X_0$ is not always as smooth as in fig. 1.1, but can sometimes show 'resonances': for some ($\alpha$, $\beta$) pairs $e_{\text{out}}$ or $e_{\text{in}}$ can rapidly increase, perhaps by as much as a factor of 4, and then decrease again in a narrow range of $X_0$, usually at $X_0 \sim 4$ or $4.5$. Surprisingly, this variation appears also to be quite smooth and systematic with $X_0$, although on the scale of fig. 1.1 it would appear as a spike. For yet other ($\alpha$, $\beta$) pairs the resonance is 'disruptive': for some central values in the peak the system breaks up, typically by ejection of $*3$, even though the system appears to be stable over very long intervals (perhaps indefinitely) at smaller as well as larger values of $X_0$, although of course the system is bound to break up at some smaller

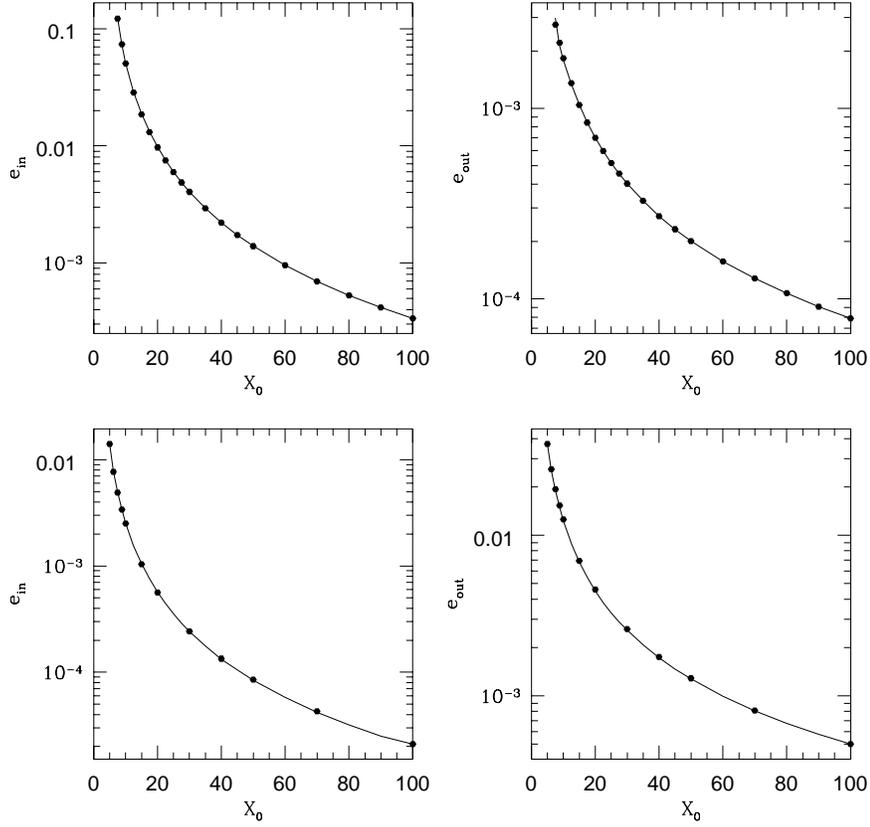

**Figure 1.1**: The mean values over time of the inner and outer eccentricities in hierarchical triples with coplanar initially circular orbits, as functions of the initial period ratio $X_0$. The inner binary contains two stars of equal mass ($\alpha=0$); the third star has $\sim 16$ (upper panels) and $\sim 1/16$ (lower panels) times the mass of the inner binary ($\beta=-1.2$ and $1.2$ respectively).

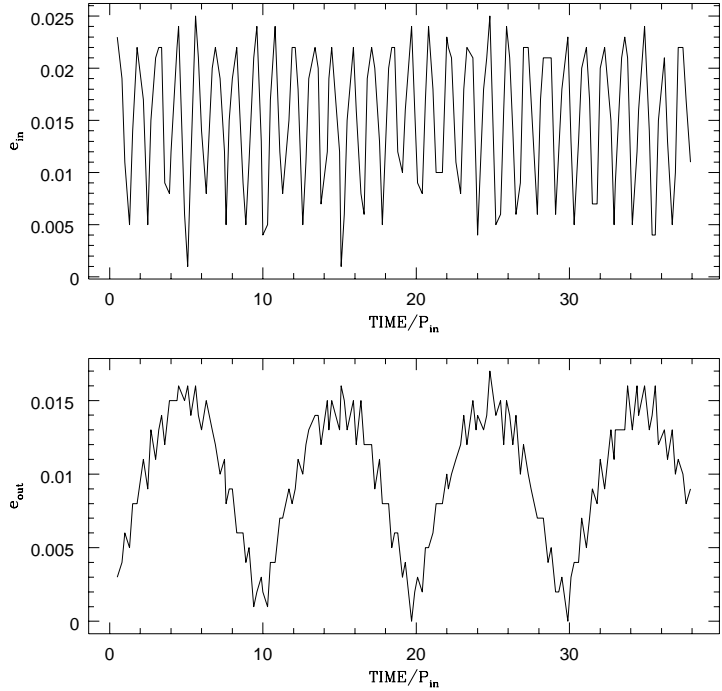

**Figure 1.2**: The fluctuations of inner and outer eccentricity with time for a system with $m_1 = m_2 = m_3$ and $X_0 = 10$. The jagged character of the curves is due to the fact that the Bulirsch-Stoer integration procedure used in our numerical simulations requires only 4 or 5 steps per inner orbit.

$X_0$ still.

It is clear that the functions in fig. 1.1 are rather smooth, i.e. non-resonant. A possible fitting function is

$$e = \frac{A}{X_0^{1.5C}\sqrt{X_0^C - B^C}}, \qquad (1.1)$$

where $e$ is the mean value of either $e_{\rm in}$ or $e_{\rm out}$, and $A$, $B$, $C$ are constant coefficients. For the examples shown in fig. 1.1 we have the following sets of fitting coefficients:

$\alpha = 0, \beta = -1.2 :\ A = 3.27, B = 5.8, C = 1$ for $\overline{e}_{\rm in}$
$\qquad\qquad\qquad A = 0.036, B = 0.716, C = 2/3$ for $\overline{e}_{\rm out}$.
$\alpha = 0, \beta = 1.2 :\ A = 0.206, B = 3.29, C = 1$ for $\overline{e}_{\rm in}$
$\qquad\qquad\qquad A = 0.224, B = 1.76, C = 2/3$ for $\overline{e}_{\rm out}$;

In each case the inner binary has equal masses, and the distant third body is large in the first case and small in the second. The curves in fig. 1.1 represent the integrations and the dots indicate the fitting with eq. 1.1. For the fitting of $\overline{e}_{\rm in}$ the value of the coefficient $B$ is close to the critical value of period ratio $X_0^{\rm min}$ for stability. Since eq. 1.1 fits very well (fig. 1.1), for rational choices of $C$, one might suppose it contains some physics; but this may be illusory since for some ($\alpha$, $\beta$) pairs we do not seem to get nearly so good a fit. We nevertheless hope in the future to find a fairly general formula to give the mean eccentricities injected into initially circular orbits by the presence of a third body.

**Table 1.1**: λ Tau - dynamical evolution for different $M_2$

| $M_1$ | $P_{\rm in}$ | $|\alpha|$ | $X_0$ | $\overline{X}$ | $\overline{e}_{\rm in}$ | $\overline{e}_{\rm out}$ |
|---|---|---|---|---|---|---|
| 1.827 | 4.00 | 0.6000 | 8.25 | $8.21 \pm 0.04$ | $0.0073 \pm 0.0033$ | $0.009 \pm 0.004$ |
| 1.456 | 6.81 | 0.7202 | 4.40 | $4.32 \pm 0.09$ | $0.0265 \pm 0.0126$ | $0.025 \pm 0.012$ |
| 1.416 | 7.29 | 0.7345 | 4.15 | $4.18 \pm 0.15$ | $0.0639 \pm 0.0296$ | $0.075 \pm 0.042$ |
| **1.413** | 7.32 | 0.7356 | 4.13 | | unstable | |
| 1.410 | 7.36 | 0.7371 | 4.11 | $4.02 \pm 0.09$ | $0.0304 \pm 0.0143$ | $0.024 \pm 0.011$ |
| 1.347 | 8.24 | 0.7610 | 3.69 | $3.60 \pm 0.11$ | $0.0409 \pm 0.0183$ | $0.030 \pm 0.015$ |
| 1.321 | 8.65 | 0.7700 | 3.49 | $3.43 \pm 0.13$ | $0.0549 \pm 0.0255$ | $0.471 \pm 0.023$ |
| **1.311** | 8.82 | 0.7739 | 3.42 | | unstable | |

The question of injected eccentricity is especially important for λ Tau. It is normal to suppose that an Algol orbit is almost exactly circular. This is partly because in such close systems one expects the orbit to have been circularised by tidal friction, and partly because even an eccentricity of say 0.01, which could probably not be ruled out by direct observation of the radial velocity curve, would be expected to lead to a very strongly non-constant mass transfer in the course of a single orbit. The pressure scale-height in a radiative atmosphere is $\sim 10^{-4}$ of the stellar radius, and so in the course of one orbit there would be an enormous variation in the pressure and density of material that was about to leave the surface of the loser. One might hope that tidal friction keeps the orbit almost circular despite the third-body perturbation. But it is difficult to see how this can happen, since the timescale on which the eccentricity is injected is basically the period of the outer orbit, whereas tidal friction is not thought to operate on so short a timescale, especially since the basically radiative envelopes of the A4 and B3 components (of comparable radius) are unlikely to be as dissipative as the convective envelopes normally considered. There is substantial disagreement on the timescale, and even the mechanism, for tidal friction (Tassoul 1995, Zahn 1992). But for the inner orbital eccentricity of 0.0073 (table 1.1) to be reduced to say $10^{-4}$ would require the dissipation timescale to be smaller than the orbital timescale by $\sim 70$, i.e. substantially less than a day. Thus we find this system to be quite difficult to understand.

We investigated the dynamical stability of the λ Tau triple system in the case that mass transfer takes place within the close binary from the less massive star (present mass $M_1 \approx 1.9\,M_\odot$) to the more massive one with $M_2 \approx 7.2\,M_\odot$. The close binary now has orbital period $P_{\rm in}$=4d. The distant third component has mass $M_3 \approx 0.7\,M_\odot$ and orbital period $P_{\rm out}$=33d around the inner binary. Thus the system has $|\alpha| \approx 0.6$, $\beta \approx 1.1$, $X_0 = 8.25$ in our notation; the stability evidently depends only on the modulus of $\alpha$. If we assume conservation of mass and angular momentum in the inner (Algol) binary then the orbital period $P_{\text{in}}^*$ for new $M_1^*$ and $M_2^*$ is

$$P_{\text{in}}^* = P_{\text{in}} \frac{M_1^3 M_2^3}{M_1^{*3} M_2^{*3}} \quad . \tag{1.2}$$

We computed the dynamical evolution of $\lambda$ Tau for a series of decreasing values of $M_1$ and increasing $M_2$ at constant $M_1 + M_2$. In other words, we increased the parameter $|\alpha|$ in our models keeping $\beta$ fixed. For each $\alpha$ we used the corresponding $X_0 = P_{\text{out}}/P_{\text{in}}^*$ from eq. 1.2, $P_{\text{out}}$ being constant. We found the situation described above of a disruptive resonance. In table 1.1 we give the values of $M_1$ (in solar units) and $X_0$ for stable and unstable cases. The corresponding average over time of the period ratio $\overline{X}$ and the inner and outer eccentricity are also given for stable models. These parameters fluctuate (we intend to discuss the fluctuations in a future paper), but for stable cases they have a well-determined average over time. The rms fluctuations of the parameters about their means are also given. Although we use a $\pm$ notation, the fluctuations are not random but fairly smooth and nearly periodic. We always start from two circular orbits, but the 3-body interaction injects a mean eccentricity into both orbits. We show the results in table 1.1 only for a few of the $(\alpha, X_0)$-sets which we considered. They are subdivided by horizontal lines in table 1.1, and correspond to (a) the range of stable models, (b) models near the disruptive resonance, and (c) the approach to the final unstable model.

We see that the present $e_{\text{in}}$ for $\lambda$ Tau should be 0.0073, and that this should increase strongly as $M_1$ continues to decrease. In the absence of the dissipative effect of tidal friction the system should break up at the disruptive resonance at $M_1 = 1.413$. However, it is not impossible that dissipation might manage to stabilise the system against disruption at this point, particularly if it is really as powerful as our above argument appears to imply, and so the system may be able to evolve further, to $M_1 = 1.311$, before being finally disrupted by the ejection of $*3$.

It is not clear, however, that the evolution of the SD pair in $\lambda$ Tau must conserve angular momentum as assumed in table 1.1, because the effect of tidal friction, even if not strong enough (i.e. on an orbital timescale) to circularise the orbit, will probably be to transfer angular momentum from the inner to the outer orbit on a long timescale (Kiseleva & Eggleton 1995), which may perhaps be comparable to the nuclear timescale. This might prevent $X_0$ from decreasing so much in the future. But we must also bear in mind the possibility that the outer orbit is *retrograde*, in which case the effect of dissipation will be to decrease the angular momenta of both orbits (moduluswise), and bring $*3$ in closer. In any event, $\lambda$ Tau is a fascinating object that combines the study of internal evolution, of dynamical evolution, and of tidal friction more closely than in any other system we are aware of. The next closest Algol triple is

DM Per   ((A6III+B5; 1.8+5.8 $M_\odot$ 2.73d) + B7:; 7.6+3.6: $M_\odot$; 100:d),

where the data are from Hilditch et al. 1986. The period ratio of 35 probably means that dynamical effects are less important here, even though ∗3 is substantially more massive, and it also means that the outer orbit is unlikely to be disrupted by the SD evolution of the inner pair. If ∗3 in λ Tau were as massive as that in DM Per, then it would not be ∗3 that is ejected when the Algol pair widens, but much more probably ∗1, since this would now be the least massive of the three components at this stage. The outcome would be a young, single, high-velocity RG, leaving behind an apparently rather normal binary of two MS stars.

For all values of $\alpha, \beta$ that we have considered, $X_0 \gtrsim 20$ ensures the preservation of the initial (circular) orbital elements of both binaries to better than 1%. Further, $X_0 \gtrsim 10$ ensures that their average changes never go beyond 5%, even when $\beta$ is negative, i.e. when the third component is more massive than the binary. The average orbital eccentricity of the inner binary depends very little on the inner mass ratio; however such dependence is stronger for the outer eccentricity.

## 1.4  Hierarchical Triple and Quadruple Stars in Clusters

Known triple systems are not so numerous in open clusters as in the field, but the statistics are increasing due to the improvement of observational techniques, and to the systematic surveys undertaken at several observatories within the last few years. There is thus growing evidence for the existence of triple and even quadruple systems in open clusters, with a variety of characteristics. These systems are usually highly hierarchical. Triple (or even higer multiplicity) systems are found in the Pleiades (Mermilliod et al. 1992), the Hyades (Griffin & Gunn 1981, Griffin et al. 1985, Mason et al. 1993), Praesepe (Mermilliod et al. 1994), M67 (Mathieu et al. 1990), and NGC 1502 (Mayer et al. 1994). The system in NGC 1502 contains an eclipsing binary SZ Cam ($13.7 + 9.7\,M_\odot$, 2.7d) which is the brightest member of the cluster. Previous studies (Chochol 1980, Mardirossian et al. 1980) showed that this binary is semidetached, but at the moment the semidetached nature of this system is questionable (Mayer et al. 1994). The variability of the orbital period of this binary has been known for some time, but only recently new high-dispersion spectra (Mayer et al. 1994) have allowed the third body to be identified. Because of its large mass (minimum $18.6\,M_\odot$) and observed shifts in the third-body lines, this 'third body' can possibly be a binary, and the system as a whole may be a hierarchical quadruple system ($P_{\rm out} = 50.7$y, $e = 0.77$).

Mermilliod et al. (1994) have summarised the data for 11 main-sequence triple systems known so far in open clusters, in which one component is a spectroscopic binary. Four of these systems contain a very close binary ($P_{\rm in} \in (2.4, 4.0)$d). Only 3 out of 11 outer orbital periods are known, and the least hierarchical system (vB 124 in the Hyades) has a period ratio $P_{\rm out}/P_{\rm in} \approx 250$.

A particularly large fraction of hierarchical triple and quadruple systems can be observed among pre-main-sequence stars in star-forming regions. For example, Ghez et al. (1993) found that triples and qudruples comprise 14% of their sample for the Tau-Aur association. They estimate that the real frequency (taking into account the incomplete period coverage in their sample) may reach $\approx 35$%. Of

*References* 17

course, some fraction of these systems may be unstable, and we observe them at the stage of the distant ejection of one companion.

Only one hierarchical triple system has been detected so far in globular clusters, but this is surely only the first step to the discovery of others which are likely to be present in these clusters. This famous system in M4 contains the millisecond pulsar PSR B1620-26 (Backer et al. 1993, Thorsett et al. 1993; see also Rasio et al. 1995 and Hut 1995 for a discussion).

The above data indicate the importance of the numerical study of the formation and evolution of hierarchical systems in star clusters. We have performed some numerical simulations for clusters of 500 - 5,000 stars with different fractions of primordial binaries, using the N-body code NBODY4 (Aarseth 1995, in preparation). This code takes into account not only dynamical interaction but also stellar evolutionary and tidal effects. The method and the first tentative results of these simulations are described in Kiseleva et al. (1995). The code identifies stable hierarchical triples when they form, using the stability criterion of Eggleton & Kiseleva (1995) and some other conditions (see Kiseleva et al. 1995), and also identifies when they are broken up (by the perturbing effects of the remaining stars or/and because of some stellar evolutionary effects in the inner binary). The main conclusion is that the formation of hierarchical systems in open clusters is rather common at almost all stages of cluster evolution. In clusters with primordial binaries the first hierarchies appear during the first $10^7$ years of cluster evolution, and in small clusters much earlier, i.e. shortly after core collapse has occured. These multiples are often destroyed later, but may nevertheless persist for some considerable time. On average at least one hierarchical system (triple or quadruple) formed by dynamical capture is present in the cluster during 10 - 20% of its lifetime, and some systems survive for more than $10^7$ years with more than $10^3$ outer revolutions. So far, we have not followed the processes of formation of hierarchical triples in clusters in detail, but a recent study of binary-binary interactions by Bacon et al. (1995) shows very good agreement with our rate of triple formation. Note also that in large clusters at a late stage of their evolution many new hierarchical systems are formed via repetitive triple-binary and triple-triple exchanges. The distributions of outer eccentricities and periods for hierarchical systems produced in our models are in reasonable agreement with the observations referred to above of multiples in open clusters, within the limited statistics for the latter.

*Acknowledgements* The authors thank Drs S. Aarseth, D. Heggie and R. de la Fuente Marcos for discussions and cooperation. LGK thanks the Isaac Newton Trust and NATO for their financial support and IoA for its hospitality.